\documentclass [12pt]{iopart}

\begin{document}
\jl{2}

\letter{Convergence improvement for coupled cluster calculations}
\author{N S Mosyagin\dag\footnote[3]{
  E-mail for correspondence: Mosyagin@lnpi.spb.su;
  http://www.qchem.pnpi.spb.ru
},
        E Eliav\ddag\ and U Kaldor\ddag}
\address{\dag\ Petersburg Nuclear Physics Institute, \\
         Gatchina, St.-Petersburg district 188350, Russia}
\address{\ddag\ School of Chemistry, Tel Aviv University,
         Tel Aviv 69978, Israel}

\begin{abstract}
Convergence problems in coupled-cluster iterations are discussed, and a new
iteration scheme is proposed. Whereas the Jacobi method inverts only the
diagonal part of the large matrix of equation coefficients, we invert a
matrix which also includes a relatively small number of off-diagonal coefficients,
selected according to the excitation amplitudes undergoing the largest change
in the coupled cluster iteration. A test case shows that
the new IPM (inversion of partial matrix) method
gives much better convergence than the straightforward Jacobi-type
scheme or such well-known convergence aids as the reduced linear equations
or direct inversion in iterative subspace methods.

\end{abstract}
\submitted
\maketitle

%
%
%

\nosections
The coupled cluster (CC) method is widely used in electronic structure
calculations.  The CC theory has been described in many reviews (see, e.g.,
\cite{bartlett,mukhrev,kaldortca,paldus,bartlett_yark}),
and will not be presented here. The basic
equation for the CC method is the Bloch equation
\begin{equation}
\mathbf{\Omega H \Omega} = \mathbf{H \Omega},
\end{equation}
where $\mathbf{H}$ is the Hamiltonian and $\mathbf{\Omega}$ is the wave
operator. The resulting equations have the general algebraic form
\begin{equation}
A_i + \sum_{j=1}^N B(t)_{ij} t_j = 0 , \qquad i=1,2,\ldots,N\;,
\label{1}
\end{equation}
where $t_j$ are the cluster or excitation amplitudes to be determined,
$N$ is the number of the unknown amplitudes, $A$ is a vector and
$B(t)$ is a square matrix which in general depends upon $t$.
For simplicity, we consider the case when $B$ does not depend upon $t$,
\begin{equation}
A_i + \sum_{j=1}^N B_{ij} t_j = 0 , \qquad i=1,2,\ldots,N.
\label{2}
\end{equation}
The generalization for the case of $B(t)$ is straightforward (and
implemented in the relativistic CC code employed for the test examples
below). The direct solution of equations~(\ref{2}) using the
Gauss elimination method is feasible only for systems with a few thousand
cluster amplitudes at most, whereas problems encountered in our relativistic
CC may involve millions of such amplitudes. A Jacobi-type iterative method
is usually applied to solve these equations. Using the fact that $B$ is
normally a diagonally dominant matrix, the method involves direct inversion
of the diagonal part $D$ of $B$. The system~(\ref{2}) is rewritten in the form
\begin{equation}
t_i= -(D^{-1})_{ii}[A_i+\sum_{j=1}^N (B-D)_{ij} t_j] , \qquad i=1,2,\ldots,N
\label{3}
\end{equation}
and is solved iteratively.

The coupled cluster calculations are often beset by convergence difficulties.
This is particularly true for multireference CC methods, such as the
Fock-space approach \cite{mukhrev,kaldortca}. Several methods for
improving convergence have been proposed; the most commonly used are the
reduced linear equations (RLE) \cite{RLE} and direct inversion in the
iterative subspace (DIIS) \cite{diis1,diis2} approaches. These help in some,
but not all, cases. Most severe convergence problems may be traced to the
existence of intruder states. While increasing the model (or $P$) space
improves the quality of the calculation by including a larger part of
the correlation, it also increases the probability of encountering intruder
states and getting no valid results at all. New methods for improving
convergence are therefore highly desirable. One such method is presented
in this Letter.

The problem may be illustrated by an example taken from recent work \cite{Hg},
where ground and excited state energies of Hg and its ions were calculated
by the relativistic coupled cluster method.
The $5d^{10}$ ground state of the Hg$^{2+}$ ion served as the reference
state, and the Fock-space CC scheme was
\begin{equation}
{\rm Hg}^{2+}{[\rm (0) sector]} \rightarrow {\rm Hg}^+{[\rm (1) sector]}
\rightarrow {\rm Hg}{[\rm (2) sector]},
\end{equation}
with electrons added in the $6s$ and $6p$ orbitals, designated as valence
particles. While
the calculations in~\cite{Hg} were relativistic, the nonrelativistic notation
will be employed for brevity. The model space in the (1) sector, with
one valence particle,
consisted of determinants with $5d^{10} 6s^1$ and $5d^{10} 6p^1$
configurations. Adding the $7s$, $7p$, and $6d$ orbitals to the list of
valence particles would yield more state energies in the (1) sector, as well as
better description of the $6s^1 6p^1$ states\cite{Hg}
in the (2) sector, corresponding to
neutral Hg. Unfortunately, adding the $5d^{10} 7s^1$, $5d^{10} 7p^1$, and
$5d^{10} 6d^1$ configurations to the model space
leads to divergence of the CC iterations~(\ref{3}) in the (1) sector.
Analysis shows that the divergence is caused by the
$5d^9 6s^1 6p^1$, $5d^9 6s^2$ and other
intruder states from the complementary $Q$ space, which are
close in energy to certain $P$ states
($5d^{10} 7p^1$, $5d^{10} 6d^1$, and others).
The diagonal elements $B_{ii}$ of the matrix $B$ correspond to
differences between the total energies of the $P$ and $Q$
determinants connected by the $t_i$
excitations. Some of these elements will be very small in this
case, leading to large elements in $D^{-1}$. Small changes
in $t$ amplitudes on the right hand side of equations~(\ref{3})
will therefore cause large changes in the amplitudes on the left hand side,
leading to divergence.

We propose to overcome this problem by replacing  the $D$ matrix
by $D'$ which includes, in addition to the diagonal elements of $B$,
those nondiagonal $B$ elements which are large in comparison with
corresponding diagonal elements.
The calculation of all $B$ matrix elements is impractical, and a selection
procedure for nondiagonal elements to be included in $D'$ is described below.
This new matrix is constructed so that its matrix
elements, $D'_{ij}$, are equal to or approximate the $B_{ij}$ matrix elements
both for $i=j$ and for $i,j\in I$, where $I$ is some small subset of
the amplitudes. The other nondiagonal $D'_{ij}$ matrix elements
($i\not\in I$ or $j\not\in I$) are set to zero. The method involves the
inversion of the partial matrix (IPM) $D'$. A modified form of
the system of equations~(\ref{3}),
\begin{equation}
t_i= -\sum_{k=1}^N (D'^{-1})_{ik}[A_k+\sum_{j=1}^N (B-D')_{kj} t_j] ,
                     \qquad i=1,2,\ldots,N
\label{5}
\end{equation}
is obtained and solved iteratively. Equations~(\ref{5}) can be divided
into two sets,
\begin{eqnarray}
\label{6}
t_i= -\sum_{k\in I} (D'^{-1})_{ik}[A_k+\sum_{j=1}^N (B-D)_{kj} t_j
                    - \sum_{j\in I} (D'-D)_{kj} t_j] , \qquad&
                      i\in I,  \\
\label{7}
t_i= -(D'^{-1})_{ii}[A_i+\sum_{j=1}^N (B-D)_{ij} t_j -(D'-D)_{ii}t_i] , &
                      i\not\in I,
\end{eqnarray}
where the second part is similar to equations~(\ref{3}).

The size of the subset $I$ must be kept small, so that the calculation
(the most time-consuming step), storage and manipulation of the non-zero
off-diagonal $D'$ elements remains feasible.
Careful selection of the amplitudes to be included in $I$ is therefore of
paramount importance. The algorithm followed here starts with calculating
the $t_i$ amplitudes by the standard iteration scheme~(\ref{3}). The
amplitudes which have undergone the largest changes are included in $I$,
the corresponding $D'_{ij}$ matrix elements are evaluated, and the $t_i$
amplitudes in $I$ are recalculated by
equations~(\ref{6}). The dimension $M$ of the $I$ subset was kept at 1000,
which makes the calculation and manipulation of $D'$ feasible.
Optimal algorithms for determining $M$, selecting excitations
to be included in $I$, and calculating the $D'$ matrix will be studied
in the future.

It should be noted that the system~(\ref{5}) is equivalent to the standard
equations~(\ref{3}) in the limit $M=0$; in the limit $M=N$,
scheme~(\ref{5}) converges in one iteration, if one takes $D'_{ij}=B_{ij}$,
\begin{equation}
t_i= -\sum_{k=1}^N (B^{-1})_{ik} A_k , \qquad i=1,2,\ldots,N .
\label{8}
\end{equation}
Formally, one can always achieve convergence of the iterations~(\ref{5})
by increasing $M$. The IPM method proposed here may be combined with
other procedures for accelerating convergence, such as
the reduced linear equations~\cite{RLE} and direct inversion in the
iterative subspace~\cite{diis1,diis2} methods. This has not been done in
the present application, and will be tried in the future.
It should be mentioned that the identification of the resulting
high-lying levels may require careful analysis of the $t$ amplitudes,
particularly if some of the latter are large, indicating large contributions
of $Q$ configurations.
Finally, it should be noted that the IPM scheme described above may be
regarded as adopting the Gershgorn-Shavitt $A_k$ perturbation theory approach
rather than that of $A_0$~\cite{Gershgorn}.

The different iteration schemes were tested for the 33-electron
relativistic Fock-space CC calculation with single and double
cluster amplitudes of Hg$^+$ levels in the $(spdfg)$ basis from~\cite{Hg}
in the framework of the Dirac-Coulomb Hamiltonian.
Two model spaces were used, one consisting of determinants with $5d^{10} 6s^1$
and $5d^{10} 6p^1$ configurations, the other including in addition the
$5d^{10} 7s^1$, $5d^{10} 7p^1$, and $5d^{10} 6d^1$ configurations.
All iterations involved 1:1 damping (the input amplitudes for iteration
$n+1$ were taken as the average of input and output amplitudes of
iteration $n$). The IPM scheme is compared with the standard scheme (\ref{3})
and with the RLE \cite{RLE} and DIIS \cite{diis1,diis2} methods in
tables \ref{tbl:small} and \ref{tbl:large}. The RLE and DIIS methods used
the output of the last five iterations to form the new input vector.
All methods led to convergence for the small model space
(table \ref{tbl:small}). The RLE, DIIS, and IPM schemes were about equally
effective in reducing the number of iterations required.
The large model space (table \ref{tbl:large}) shows markedly
different behavior for the different methods. Straightforward iteration by
the Jacobi-type method blows up almost immediately; the large excitation
amplitudes may be traced to the intruder states mentioned above. The RLE
and DIIS schemes exhibit better behavior, but could not achieve convergence
even after several hundred iterations. Only the IPM approach proposed in this
Letter led to convergence (in the 29th iteration), showing the potential
of the method.

\ack

This work was supported by INTAS grant No 96--1266. N~M thanks the
Russian Foundation for Basic Research (grant No 99--03--33249).
Work at TAU was supported by the Israel Science Foundation.
The authors are grateful for valuable discussions with A.V.~Titov.

\Tables

\begin{table}
\caption{ The largest change in the single and double cluster amplitudes
          ($\displaystyle \max_{i=1}^N|t_i^{(n+1)}-t_i^{(n)}|$)
	  at iteration $n$.
	  The changes are obtained by equations~(\ref{3})
	  in the RCC calculations with the Jacobi-type,
	  RLE, DIIS and IPM iteration schemes.
	  The model space consists of determinants with $5d^{10} 6s^1$ and
	  $5d^{10} 6p^1$ configurations.
	  The convergence threshold is $10^{-6}$.
          }
\label{tbl:small}
\begin{tabular}{@{}ccccc}
\br
 Iteration  & Jacobi            & RLE                & DIIS               & IPM               \\
\mr
 ~~0        &$1.62\cdot 10^{-1}$&$1.62\cdot 10^{-1}$ &$1.62\cdot 10^{-1} $&$1.62\cdot 10^{-1}$\\
 ~~3        &$2.99\cdot 10^{-2}$&$2.99\cdot 10^{-2}$ &$2.59\cdot 10^{-2} $&$1.30\cdot 10^{-2}$\\
 ~~6        &$1.39\cdot 10^{-2}$&$1.41\cdot 10^{-3}$ &$1.10\cdot 10^{-3} $&$1.38\cdot 10^{-3}$\\
 ~~9        &$6.65\cdot 10^{-3}$&$6.81\cdot 10^{-4}$ &$1.21\cdot 10^{-4} $&$2.01\cdot 10^{-4}$\\
  12        &$3.19\cdot 10^{-3}$&$1.69\cdot 10^{-5}$ &$1.12\cdot 10^{-5} $&$3.60\cdot 10^{-5}$\\
  15        &$1.54\cdot 10^{-3}$&$8.42\cdot 10^{-6}$ &$5.15\cdot 10^{-6} $&$7.01\cdot 10^{-6}$\\
  18        &$7.53\cdot 10^{-4}$&$9.47\cdot 10^{-7}$ &$2.43\cdot 10^{-6} $&$1.43\cdot 10^{-6}$\\
  21        &$3.72\cdot 10^{-4}$& convergence        &$1.15\cdot 10^{-6} $& convergence       \\
  24        &$1.86\cdot 10^{-4}$&                    &  convergence       &                   \\
  27        &$9.46\cdot 10^{-5}$&                    &                    &                   \\
  30        &$4.87\cdot 10^{-5}$&                    &                    &                   \\
  33        &$2.53\cdot 10^{-5}$&                    &                    &                   \\
  36        &$1.34\cdot 10^{-5}$&                    &                    &                   \\
  39        &$7.11\cdot 10^{-6}$&                    &                    &                   \\
  42        &$3.82\cdot 10^{-6}$&                    &                    &                   \\
  45        &$2.07\cdot 10^{-6}$&                    &                    &                   \\
  48        &$1.13\cdot 10^{-6}$&                    &                    &                   \\
            & convergence       &                    &                    &                   \\
\br
\end{tabular}
\end{table}

\begin{table}
\caption{ Same as Table 1, except that the model space is larger,
	  consisting of determinants with $5d^{10} 6s^1$,
	  $5d^{10} 6p^1$, $5d^{10} 7s^1$, $5d^{10} 7p^1$, and
          $5d^{10} 6d^1$ configurations.
          }
\label{tbl:large}
\begin{tabular}{@{}ccccc}
\br
 Iteration  & Jacobi            & RLE                & DIIS               & IPM               \\
\mr
 ~~0        &$2.14\cdot 10^{-1}$&$2.14\cdot 10^{-1}$ &$2.14\cdot 10^{-1} $&$2.14\cdot 10^{-1}$\\
 ~~3        &$7.54\cdot 10^{-1}$&$7.54\cdot 10^{-1}$ &$6.93\cdot 10^{-1} $&$9.38\cdot 10^{-2}$\\
 ~~6        &$2.61~~~~~~~~     $&$5.18\cdot 10^{-1}$ &$1.22\cdot 10^{-1} $&$1.21\cdot 10^{-2}$\\
 ~~9        &$9.31~~~~~~~~     $&$1.82~~~~~~~~     $ &$5.75\cdot 10^{-2} $&$1.25\cdot 10^{-3}$\\
  12        &$3.42\cdot 10^{1~~}$&$2.73\cdot 10^{-1}$ &$3.20\cdot 10^{-2}$&$3.19\cdot 10^{-4}$\\
  15        &$1.24\cdot 10^{2~~}$&$5.65\cdot 10^{-1}$ &$2.90\cdot 10^{-2}$&$7.39\cdot 10^{-5}$\\
  18        &$4.45\cdot 10^{2~~}$&$1.80\cdot 10^{1~~}$ &$2.87\cdot 10^{-2}$&$1.61\cdot 10^{-5}$\\
  21        &$1.54\cdot 10^{3~~}$&$2.77\cdot 10^{-1}$ &$3.06\cdot 10^{-2}$&$1.04\cdot 10^{-5}$\\
  24        &$5.00\cdot 10^{3~~}$&$1.01~~~~~~~~     $ &$2.55\cdot 10^{-2}$&$9.55\cdot 10^{-6}$\\
  27        &$1.49\cdot 10^{4~~}$&$1.46\cdot 10^{-1}$ &$2.19\cdot 10^{-2}$&$3.64\cdot 10^{-6}$\\
  30        &$3.93\cdot 10^{4~~}$&$5.51\cdot 10^{-1}$ &$2.23\cdot 10^{-2}$& convergence       \\
  33        &$9.08\cdot 10^{4~~}$&$2.40\cdot 10^{-1}$ &$1.93\cdot 10^{-2}$&                   \\
  36        &$1.88\cdot 10^{5~~}$&$1.73\cdot 10^{-1}$ &$1.99\cdot 10^{-2}$&                   \\
  39        &$3.61\cdot 10^{5~~}$&$4.91\cdot 10^{-1}$ &$1.36\cdot 10^{-2}$&                   \\
  42        &$6.70\cdot 10^{5~~}$&$2.95\cdot 10^{-1}$ &$1.62\cdot 10^{-2}$&                   \\
  45        &$1.22\cdot 10^{6~~}$&$6.04\cdot 10^{-1}$ &$1.44\cdot 10^{-2}$&                   \\
  48        &$2.22\cdot 10^{6~~}$&$5.00\cdot 10^{-1}$ &$1.45\cdot 10^{-2}$&                   \\
            & divergence         &  no convergence    &no convergence     &                   \\
\br
\end{tabular}
\end{table}

\end{document}